%%%%%%%%%%%%%%%%%%%%%%%%%%%%%%%%%%%%%%%%%%%%%%%%%%%%%
%  TeX format and instructions follow
%%%%%%%%%%%%%%%%%%%%%%%%%%%%%%%%%%%%%%%%%%%%%%%%%%%%%

%****************************************************
%bcp.tex
%****************************************************

\frenchspacing

\parindent15pt

\abovedisplayskip4pt plus2pt
\belowdisplayskip4pt plus2pt 
\abovedisplayshortskip2pt plus2pt 
\belowdisplayshortskip2pt plus2pt  

\font\twbf=cmbx10 at12pt
 at12pt
 at12pt

\font\sc=cmcsc10

\font\ninerm=cmr9 
\font\nineit=cmti9 
\font\ninesy=cmsy9 
\font\ninei=cmmi9 
\font\ninebf=cmbx9 

\font\sevenrm=cmr7  
 
\font\seveni=cmmi7  
\font\sevensy=cmsy7 

\font\fivenrm=cmr5  
\font\fiveni=cmmi5  
\font\fivensy=cmsy5 

\def\nine{%
\textfont0=\ninerm \scriptfont0=\sevenrm \scriptscriptfont0=\fivenrm
\textfont1=\ninei \scriptfont1=\seveni \scriptscriptfont1=\fiveni
\textfont2=\ninesy \scriptfont2=\sevensy \scriptscriptfont2=\fivensy
\textfont3=\tenex \scriptfont3=\tenex \scriptscriptfont3=\tenex
\def\rm{\fam0\ninerm}%
\textfont\itfam=\nineit    
\def\it{\fam\itfam\nineit}%
\textfont\bffam=\ninebf 
\def\bf{\fam\bffam\ninebf}%
\normalbaselineskip=11pt
\setbox\strutbox=\hbox{\vrule height8pt depth3pt width0pt}%
\normalbaselines\rm}

\hsize30cc
\vsize44cc
\nopagenumbers

\def\luz#1{\luzno#1?}
\def\luzno#1{\ifx#1?\let\next=\relax\yyy
\else \let\next=\luzno#1\xxx\fi\next}
\def\sp#1{\def\xxx{\kern1.7pt}\def\yyy{\kern-1.7pt}\luz{#1}}
\def\spa#1{\def\xxx{\kern1pt}\def\yyy{\kern-1pt}\luz{#1}}

\newcount\beg
\newbox\aabox
\newbox\atbox
\newbox\fpbox
\def\abbrevauthors#1{\setbox\aabox=\hbox{\sevenrm\uppercase{#1}}}
\def\abbrevtitle#1{\setbox\atbox=\hbox{\sevenrm\uppercase{#1}}}
\long\def\pag{\beg=\pageno
\def\leftheadline{\noindent\rlap{\nine\folio}\hfil\copy\aabox\hfil}
\def\rightheadline{\noindent\hfill\copy\atbox\hfill\llap{\nine\folio}}
\def\phead{\setbox\fpbox=\hbox{\sevenrm 
************************************************}%
\noindent\vbox{\sevenrm\baselineskip9pt\hsize\wd\fpbox%
\centerline{***********************************************}

\centerline{BANACH CENTER PUBLICATIONS, VOLUME **}

\centerline{INSTITUTE OF MATHEMATICS}

\centerline{POLISH ACADEMY OF SCIENCES}

\centerline{WARSZAWA 19**}}\hfill}
\footline{\ifnum\beg=\pageno \hfill\nine[\folio]\hfill\fi}
\headline{\ifnum\beg=\pageno\phead
\else
\ifodd\pageno\rightheadline \else \leftheadline \fi 
\fi}}

\newbox\tbox
\newbox\aubox
\newbox\adbox
\newbox\mathbox

\def\title#1{\setbox\tbox=\hbox{\let\\=\cr 
\baselineskip14pt\vbox{\twbf\tabskip 0pt plus15cc
\halign to\hsize{\hfil\ignorespaces \uppercase{##}\hfil\cr#1\cr}}}}

\newbox\abbox
\setbox\abbox=\vbox{\vglue18pt}

\def\author#1{\setbox\aubox=\hbox{\let\\=\cr 
\nine\baselineskip12pt\vbox{\tabskip 0pt plus15cc
\halign to\hsize{\hfil\ignorespaces \uppercase{\spa{##}}\hfil\cr#1\cr}}}%
\global\setbox\abbox=\vbox{\unvbox\abbox\box\aubox\vskip8pt}}

\def\address#1{\setbox\adbox=\hbox{\let\\=\cr 
\nine\baselineskip12pt\vbox{\it\tabskip 0pt plus15cc
\halign to\hsize{\hfil\ignorespaces {##}\hfil\cr#1\cr}}}%
\global\setbox\abbox=\vbox{\unvbox\abbox\box\adbox\vskip16pt}}

\def\mathclass#1{\setbox\mathbox=\hbox{\footnote{}{1991 {\it Mathematics Subject 
Classification}\/: #1}}}

\long\def\maketitlebcp{\pag\unhbox\mathbox
\footnote{}{The paper is in final form and no version 
of it will be published elsewhere.} 
\vglue7cc
\box\tbox
\box\abbox
\vskip8pt}

\long\def\abstract#1{{\nine{\bf Abstract.} 
#1

}}

\def\section#1{\vskip-\lastskip\vskip12pt plus2pt minus2pt
{\bf #1}}

\long\def\defin#1#2{\vskip-\lastskip\vskip4pt plus2pt
{\sc #1} #2 \vskip-\lastskip\vskip4pt plus2pt}

\long\def\remar#1#2{\vskip-\lastskip\vskip4pt plus2pt
\sp{#1} #2\vskip-\lastskip\vskip4pt plus2pt}

\def\endproof{\nobreak\kern5pt\nobreak\vrule height4pt width4pt depth0pt
\vskip4pt plus2pt}

\newbox\refbox
\newdimen\refwidth
\long\def\references#1#2{{\nine
\setbox\refbox=\hbox{\nine[#1]}\refwidth\wd\refbox\advance\refwidth by 12pt%
\def\textindent##1{\indent\llap{##1\hskip12pt}\ignorespaces}% PLAIN
\vskip24pt plus4pt minus4pt
\centerline{\bf References}
\vskip12pt plus2pt minus2pt
\parindent=\refwidth
#2

}}

\def\footnoterule{\kern -3pt \hrule width 4cc \kern 2.6pt}

\catcode`@=11
\def\vfootnote#1%
{\insert\footins\bgroup\nine\interlinepenalty\interfootnotelinepenalty%
\splittopskip\ht\strutbox\splitmaxdepth\dp\strutbox\floatingpenalty\@MM%
\leftskip\z@skip\rightskip\z@skip\spaceskip\z@skip\xspaceskip\z@skip%
\textindent{#1}\footstrut\futurelet\next\fo@t}
\catcode`@=12

\def\h{{\cal H}}

\def\be{\begin{equation}}
\def\ee{\end{equation}}
\def\ba{\begin{eqnarray}}
\def\ea{\end{eqnarray}}

\def\C{{\cal C}}

\def\B{\agb}

\def\Comp{{\mathchoice
{\setbox0=\hbox{$\displaystyle\rm C$}\hbox{\hbox to0pt
{\kern0.4\wd0\vrule height0.9\ht0\hss}\box0}}
{\setbox0=\hbox{$\textstyle\rm C$}\hbox{\hbox to0pt
{\kern0.4\wd0\vrule height0.9\ht0\hss}\box0}}
{\setbox0=\hbox{$\scriptstyle\rm C$}\hbox{\hbox to0pt
{\kern0.4\wd0\vrule height0.9\ht0\hss}\box0}}
{\setbox0=\hbox{$\scriptscriptstyle\rm C$}\hbox{\hbox to0pt
{\kern0.4\wd0\vrule height0.9\ht0\hss}\box0}}}}
\def\Co{{\mathchoice
{\setbox0=\hbox{$\displaystyle\rm C$}\hbox{\hbox to0pt
{\kern0.4\wd0\vrule height0.9\ht0\hss}\box0}}
{\setbox0=\hbox{$\textstyle\rm C$}\hbox{\hbox to0pt
{\kern0.4\wd0\vrule height0.9\ht0\hss}\box0}}
{\setbox0=\hbox{$\scriptstyle\rm C$}\hbox{\hbox to0pt
{\kern0.4\wd0\vrule height0.9\ht0\hss}\box0}}
{\setbox0=\hbox{$\scriptscriptstyle\rm C$}\hbox{\hbox to0pt
{\kern0.4\wd0\vrule height0.9\ht0\hss}\box0}}}}
\def\Rl{{\mathchoice
{\setbox0=\hbox{$\displaystyle\rm R$}\hbox{\hbox to0pt
{\kern0.4\wd0\vrule height0.9\ht0\hss}\box0}}
{\setbox0=\hbox{$\textstyle\rm R$}\hbox{\hbox to0pt
{\kern0.4\wd0\vrule height0.9\ht0\hss}\box0}}
{\setbox0=\hbox{$\scriptstyle\rm R$}\hbox{\hbox to0pt
{\kern0.4\wd0\vrule height0.9\ht0\hss}\box0}}
{\setbox0=\hbox{$\scriptscriptstyle\rm R$}\hbox{\hbox to0pt
{\kern0.4\wd0\vrule height0.9\ht0\hss}\box0}}}}

\def\C{{\cal C}}

\def\wh{\widehat}

\def\H{{\cal H}}
\def\B{{\cal B}}

\def\Ga{\Gamma}

\def\haux{\H_{aux}}
\def\r{\rangle}
\def\l{\langle}
\def\rp{\rangle_{phys}}

\def\hp{\H_{phys}}
\def\V{{\cal V}}
\def\vp{\V_{phys}}

\def\bst{{\cal B}^{(\star)}}
\def\baux{{\cal B}_{aux}}
\def\bsa{\bst_{aux}}
\def\bsp{\bst_{phys}}
\def\Ps{\Phi'}

\def\en{\eta^{[\tilde \alpha]}}

\def\C*{$C^{\star}$}
\def\ge{\geq}

% {${}\qquad\spadesuit$}

% preprint UCSBTH-95-16

\mathclass{Primary 81V17; Secondary 81Q10.}

\abbrevauthors{D. Marolf}
\abbrevtitle{Quantizing a single Constraint}

\title{Refined Algebraic Quantization:\\  Systems with a single
constraint}

\author{Donald\ Marolf}
\address{Physics Department, University of California\\
Santa Barbara, California 93106 USA\\
E-mail: marolf@cosmic.physics.ucsb.edu}

\maketitlebcp

\footnote{}{This work was supported by NSF grant PHY90-08502.}

\abstract{
This paper explores in some detail
a recent proposal (the Rieffel induction/refined
algebraic quantization scheme) for the quantization of
constrained gauge systems.  Below, the focus is on
systems with a single constraint and, in this context, on
the uniqueness of the construction.  While in general the
results depend heavily on the choices made for certain auxiliary
structures, an additional physical argument leads to a unique result for
typical cases.  We also discuss the `superselection laws'
that result from this scheme and how their existence also depends on
the choice of auxiliary structures.  Again, when these structures
are chosen in a physically motivated way, the resulting superselection
laws are physically reasonable.}

\section{1. Introduction}

Canonical quantization of gauge systems has been a subject of much
discussion since the basic outline was first given by Dirac [5].
This formalism has been especially popular in the gravitational
physics community as, for Einstein's general theory of relativity on a
spatially compact universe, the Hamiltonian consists {\it only}
of constraints.  In addition, the nature of the gauge transformations
associated with gravity
make gauge fixing techniques extremely difficult to apply and 
perturbative nonrenormalizability has frustrated  attempts at covariant
path integral quantization.  Thus, Dirac style canonical quantization 
remains at the forefront of quantum gravity research [2,10,11].

Despite this interest,  certain basic issues
have remained unresolved for the general case.  Recall that
the essential idea of Dirac's
approach is to turn the classical constraints $C_i$
into linear operators $\wh{C}_i$ and to consider `physical
states $\psi_{phys}$' that are annihilated by the constraints; i.e., such
that $\wh{C}_i \psi_{phys} = 0$.
However, the questions 
of on which linear space the 
constraints should act and of just how an inner product is to be
imposed on the solutions to define a Hilbert space do not yet have
widely accepted answers.

Recently, a resolution to these issues has been proposed.
In fact, what is 
essentially the same resolution has been
independently
suggested twice under the names of the `Rieffel induction procedure'
[12] and the `refined algebraic quantization scheme' [1].
This method has been successfully used to quantize linearized
gravity on symmetric backgrounds [8,9], minisuperspace models for
gravity [16,17], and the free Maxwell field [13].
As might be expected, these methods proceed by introducing 
additional structures beyond what is present in the original
Dirac approach.  These techniques  and additional structures
will be explored further here in the particular context of 
systems with a single constraint.  

We begin with a review of the Rieffel/refined algebraic procedure in
section 2.  Here, we use the language and notation of [1]
as it is more closely related to that of Dirac [5] and therefore more
familiar.  We will also refer to the scheme as the `refined
algebraic proposal' in the text.  

Sections 3 and 4 contain the main results of this paper.  In section 3
we show how, for typical systems with a single constraint, 
a physical argument determines a unique implementation of
the Rieffel/refined algebraic scheme.  In section 4
we discuss superselection laws on the physical Hilbert space and how their 
existence may depend on the choice of auxiliary structures.  We
give two examples.  In the first, the use of an `incorrect' structure
leads to spurious superselection laws.  This example also illustrates
the fact that the physical Hilbert space can depend
strongly on the choice of this structure.
In the second, the use
of a physically motivated auxiliary structure produces superselection 
laws, but this time a similar feature exists in the classical theory.
Thus, for this second case we take the superselection laws to be
physically meaningful.  Appendices A and B contain proofs of technical 
results which are not of direct relevance to the main discussion but
which are mentioned in the text.

\section{2. The Refined Algebraic Approach} 

In this section, we review the refined algebraic quantization scheme presented
in [1] (which is essentially equivalent to the Rieffel induction
procedure of [12]) for systems with gauge symmetries.
The starting point is a constrained classical system with phase space $\Gamma$
and, as usual, the nondegenerate symplectic form $\omega$ on
$\Gamma$ defines a Poisson Bracket on smooth functions $\Gamma \rightarrow
\Co$.  The constraints $C_i$ are required to be first class; that is, the
Poisson bracket of two constraints is a sum of constraints
(possibly weighted by smooth phase space functions), as is the Poisson
bracket of any constraint with the Hamiltonian.  As a result, the 
constraint surface is preserved under time evolution.

The refined algebraic proposal quantizes this system in a series of steps 
based on those of the original algebraic quantization program [2,3].
The first four steps below have nothing to do with constrained
systems but simply quantize the system obtained by ignoring the
constraints.  They follow the unconstrained prescription of
[2,3] exactly but we repeat them
here for completeness and to fix our notation.

\vskip4pt plus2pt

\item {Step 1.} Select a subspace ${\bf S}$ of the vector space of
all smooth, complex-valued functions on $\Ga$ subject to the 
following conditions: 

\vskip4pt plus2pt

\item \item a) $\bf S$ should be large enough so that any sufficiently
regular function on the phase space can be obtained as
(possibly a suitable limit of) a sum of products
of elements in $\bf S$.

\vskip4pt plus2pt

\item 	\item b) $\bf S$ should be closed under Poisson brackets,
i.e. for all functions $F, G$ in $\bf S$, their Poisson bracket
$\{F, G \}$ should also be an element of $\bf S$.

\vskip4pt plus2pt

\item \item c) 
Finally, $\bf S$ should be closed under complex conjugation; i.e.
for all $F$ in $\bf S$, the complex conjugate $F^*$ should be
a function in $\bf S$. 

\vskip4pt plus2pt

The idea is that each function in $\bf S$ is to be regarded
as an {\it elementary classical variable} which is to have
an {\it unambiguous} quantum analog. 

\vskip4pt plus2pt

\item {Step 2.} Associate with each element $F$ in
$\bf S$ an abstract operator $\wh F$. Construct the
free associative algebra generated by these {\it elementary quantum
operators}. Impose on it the canonical commutation relations,
$[\wh F, \wh G] = i \hbar \wh{ \{ F, G \} }$, and, if necessary,
also a set of (anti-commutation) relations that captures the algebraic
identities satisfied by the elementary classical variables.
Denote the resulting algebra by $\B_{aux}$.  

\vskip4pt plus2pt

\item {Step 3.} On this algebra, introduce an involution\footnote{${}^1$}{
Recall that an involution on $\baux$ is an anti-linear map $\star$ from
$\baux$ to itself satisfying the following three conditions for all
$A$ and $ B$ in $\baux$: i) $(A + \lambda B)^\star = A^\star +
\lambda^* B^\star$, where $\lambda$ is any complex number; ii)
$(AB)^\star = B^\star A^\star$; and iii) $(A^\star)^\star = A$.
} operation
$\star$ by requiring that if two elementary classical variables $F$
and $G$ are related by $F^* = G$, then $\wh F^\star = \wh G$ in
$\baux$. Denote the resulting $\star$-algebra by ${\cal
B}^{(\star)}_{aux}$.

\vskip4pt plus2pt

\item {Step 4.}  Construct a linear $\star$-representation $R$ of
the abstract algebra $\bsa$ via linear operators on an
auxiliary Hilbert space $\haux$, i.e. such that
$$
R(\wh A^\star) =  R(\wh{A})^\dagger
$$
for all $\wh {A}$ in $\bst$, where $\dagger$ denotes Hermitian conjugation
with respect to the inner product in $\haux$.  

\vskip4pt plus2pt

The remaining steps introduce the constraints and address the
questions raised in the introduction.  That is, they first
use the space $\haux$ to provide a home for the constraints and for the linear
space on which they act, and then construct the physical Hilbert space
from the corresponding solutions.

\vskip4pt plus2pt

\item {Step 5a.} Represent the constraints $C_i$ as self-adjoint 
operators $\wh C_i$ (or, their exponentiated action, representing the
finite gauge transformations, as unitary operators $\wh U_i$) on
$\haux$.

\vskip4pt plus2pt

We will now look for solutions
of the constraints in terms of {\it generalized} eigenvectors of $\wh
C_i$ which will lie in the {\it topological dual} $\Phi'$ of some dense
subspace $\Phi \subset \haux$ (see also Ref. [6,7]).  Since $\Phi$
and $\Phi'$ will be used to build the physical Hilbert space, we will
consider only physical operators that are well behaved with respect to
$\Phi$.

\vskip4pt plus2pt

\item {Step 5b.} Choose a suitable dense subspace $\Phi \subset
\haux$ which is left invariant by the constraints $\wh{C_i}$ and let
$\bsp$ be the $\star$-algebra of operators on $\haux$ which commute
with the constraints $\wh{C}_i$ and such that, for $A \in \bsp$, both
$A$ and $A^\dagger$ are defined on $\Phi$ and map $\Phi$ to itself.

\vskip4pt plus2pt

As an example in section 4 will illustrate, some physical input is in
general required to choose the space $\Phi$.  Some factors governing this
choice are that
it must be sufficiently large so that $\bsp$ contains
`enough' physically interesting operators while it must also be
be sufficiently small that its topological dual $\Phi'$
contains enough physical states.

The main idea in the last few steps is that, while
the classical reality conditions
should determine the inner product, we should not need to explicitly
display a complete set of classical observables (i.e., functions which
Poisson commute with the constraints) to achieve this goal.  
Instead, we use a
complete set of functions (${\bf S}$) on the {\it unconstrained}
phase space, noting that the reality properties of such functions will
determine the reality properties of the observables.  
The reality conditions of operators in $\bsa$ are then implemented on the 
auxiliary Hilbert space $\haux$.  The physical Hilbert space
$\hp$ is to be constructed in such a way that any adjointness
relations involving only observables (i.e., $A = B^\dagger$, for $A,B$
observables) will in turn 
descend from $\haux$ to $\hp$ (so that $A = B^\dagger$ on
$\hp$ as well).  In this way, we will say that the
reality conditions are implemented on $\hp$.

We now wish to construct the physical Hilbert space $\hp$, which
will in general {\it not} be a subspace of $\haux$.  The
key idea is to find an appropriate map $\eta: \Phi \rightarrow
\Phi'$ such that $\eta(\phi)$ is a solution of the constraints for all
$\phi \in \Phi$. (Note that the natural class of maps from $\Phi$ to
$\Phi'$ is {\it anti}-linear (e.g., the adjoint map)).  
We proceed as follows.  

\vskip4pt plus2pt

\item {Step 5c.}  Find an anti-linear map $\eta$ from $\Phi$ to the
topological dual $\Ps$ that satisfies:

\vskip4pt plus2pt

\item \item (i) For every $\phi_1 \in \Phi$, $\eta(\phi_1)$ is a solution
of the constraints; i.e.,

$$
0 = \bigl(\wh{C}_i (\eta \phi_1)\bigr)[\phi_2] := (\eta \phi_1) [
\wh{C}_i \phi_2]
$$
for any $\phi_2 \in \Phi$.
Here, the square brackets denote the natural action of $\Ps$ on
$\Phi$.

\vskip4pt plus2pt

\item \item (ii) $\eta$ is  real and positive in the sense that, for all
$\phi_1,\phi_2 \in \Phi$, 
$$
({\eta \phi_1})[\phi_2] = ((\eta \phi_2)[\phi_1])^* {\ \ \  {\rm and}
\ \ \ } 
$$
$$
(\eta \phi_1)[\phi_1] \ge 0 $$.

\vskip4pt plus2pt

\item \item (iii) $\eta$ commutes with the action of any $A \in \bsp$ in the
sense that 
$$
(\eta \phi_1)[A \phi_2] = ((\eta A^\dagger \phi_1))[\phi_2]
$$
for all $\phi_1,\phi_2 \in \Phi$.  The r.h.s. defines the so-called
dual action of $A$ on $\Phi'$ so that we may write this as
$\eta A \phi = A \eta \phi$.

\vskip4pt plus2pt

In analogy with [12] we call $\eta$ the {\it rigging map}.
(The appearance of the adjoint on the r.h.{s.} of the above equation
corresponds to the anti-linearity of $\eta$.)

\vskip4pt plus2pt

\item {Step 5d.} The vectors $\eta \phi$ span a space $\vp$ of
solutions of the constraints. We introduce an inner product on $\vp$
through
$$
\l \eta \phi_1, \eta \phi_2 \rp = (\eta \phi_2) [ \phi_1]
$$
The requirement (iii) guarantees that this inner product is well
defined and that it is Hermitian and positive definite so that the
corresponding completion of $\vp$ is a
`physical' Hilbert space $\hp$.
(Note that the positions of $\phi_1$ and $\phi_2$ must be 
opposite on the two sides of this definition due to the anti-linear nature
of $\eta$.)

\vskip4pt plus2pt

At this point, the reader may fear that this list of conditions on
$\eta$ will never be met in practice.  That the new step 5 may
actually simplify the quantization program follows from the
observation of [8,9] (and [15,16] for the case when the
Poisson algebra of constraints is Abelian) that natural candidates
for such a map exist.  

The last step is to represent
physical operators on $\vp$. This is straightforward because the
framework provided by step 5 guarantees that $\hp$ carries an (anti)
$\star$-representation (see below) of $\bsp$ as follows:

\vskip4pt plus2pt

\item {Step 6.}  Operators in $ A \in \bsp$ have a natural action
(induced by duality) on $\Ps$ that leaves $\vp$ invariant. Use this
fact to induce densely defined operators $A_{phys}$ on $\hp$ through
$$
A_{phys}\ (\eta \phi) = \eta ( A \phi).
$$

\vskip4pt plus2pt

This leads to an {\it anti}- $\star$-representation of $\bsp$ on
$\hp$ as
the map $A \mapsto A_{phys}$ from $\bsp$ to the operators on
$\hp$ is an anti-linear $\star$-homomorphism where $\star$ acts on the
operator $A_{phys}$ in the sense of conjugation of
quadratic forms on the dense domain $\Phi$ 
($\langle \phi, A^\star \psi \rangle \equiv \langle \psi,
A \phi \rangle^*$).
In this way, the reality
properties of the physical operators $\bsp$ on $\haux$ descend to the
physical Hilbert space.

In addition, consider any $C*$ algebra with unit ${\cal B}^{C*}$ which
is a subalgebra of $\bsp$.
Since the physical expectation value $\eta(A\phi)[\phi]$ 
defines a positive functional on ${\cal B}^{C*}$ (i.e., 
 $\eta(A^\dagger A\phi)[\phi] \ge 0$), it follows that for $A
\in {\cal B}^{C*}$ we have
$$
\eta(A^\dagger A\phi)[\phi]  \le ||A||^2 
 \eta(\phi)[\phi]
$$
so that $A_{phys}$ is a bounded operator on $\hp$ with norm
not larger than that of $A$ on $\haux$ ($||A||_{phys} \le ||A||$).
Thus, for such bounded operators, any relations of the form 
$A =  B^\dagger$ on $\haux$ also hold as the adjointness relations
$A_{phys} = B^\dagger_{phys}$ on $\hp$. From this it
follows that if $A$ is self-adjoint on $\haux$ and if a sufficiently
large class of bounded functionals of $A$ map $\Phi$ to itself, then $\bsp$
determines a (unique) self-adjoint extension of $A_{phys}$
on $\hp$.

Let us consider for a moment the case where there is only one
constraint.  Note that when this constraint 
has purely discrete spectrum, there is a natural choice for the map
$\eta$ as follows.  Let $\Pi_0$ be the projection onto the eigenspace of the
constraint with eigenvalue zero.  Then if we take $\Phi =
\haux$, the rigging map $\eta$ given by 
$$
\eta |\psi \r = \l \psi | \Pi_0
$$
fulfills all the requirements of step 5c.  This case is simple and
easy to deal with, so that we shall focus on the complimentary
case (where the spectrum
is purely continuous) in the next section.  Section 4 will describe
what happens when both continuous and discrete spectra are present.

\section{3.  A Unique Prescription}

While the framework described in
section 2 sets the stage for quantizing constrained
systems, it does not provide the complete script.  There are in fact
three inputs that need to be provided in order to proceed.
The first is the auxiliary space $\haux$ itself, but the dense subspace
$\Phi \subset \haux$ and the rigging map $\eta: \Phi \rightarrow \Phi'$
must also be given.   As such, it is natural to ask to what extent
the above prescription is unique and to what extent it depends on
the choice of these inputs.  In general, the answer is that the final
formulation depends a great deal on the inputs, as different
choices can even lead to physical Hilbert spaces of different dimensions!
This will be illustrated by an example in the next section.

Below, we confine ourselves to the case of a single
constraint $\wh C$ of
the typical kind that arises in finite dimensional models. 
The two main types of constrained systems are the `classic'
gauge systems in which the constraint is a vector field (whose orbits
are closed subsets of the configuration space) on some
configuration space and the `time reparametrization invariant
systems' in which the constraint is essentially the same as some
Hamiltonian of nonrelativistic quantum mechanics (but typically
with both positive and negative kinetic terms).
For such cases, physical reasoning will lead to a preferred choice of
the dense subspace $\Phi$ such that the rigging map 
is then {\it unique} up to scale.  As we consider constraints
with continuous spectrum, we shall assume that the configuration space
is noncompact.  We argue as follows.

An important element of classical symplectic mechanics is that
the algebra of observables is taken to be the set of 
{\it smooth} functions on
the phase space (as in step 1 of the refined algebraic program).  
It is this definition, for example, that allows
us to talk about the (local) symplectomorphism `generated by an observable A.'
As such, the topology and differential structure
of the phase space play a key role and we would like to encode them
in our quantum formulation.  Consider the
case where the classical phase space is $T^* \Rl^n$ and the auxiliary
Hilbert space used in the refined quantization program is $L^2(\Rl^n)$.
Recall that one characterization of the Schwarz space ${\cal S} \subset
\haux$ is as the set of all states $|\psi \r$ for which both $\l x| \psi \r$
and $\l p| \psi \r$ are smooth $L^2$ functions of $x$ and $p$, 
where $\l x|$ and $\l p |$ are the usual position and momentum generalized 
eigenstates.  Thus, this set of states can be said to encode the
differential structure of the classical phase space and
is a natural choice for the subspace $\Phi$ of step 5b.
The algebra $\bsp$ of operators that
preserve this space contains all suitably
smooth and rapidly decreasing combinations of $x$ and $p$, in good
analogy with the classical algebra of observables.  Thus, we take
$\Phi = {\cal S}$.

We will now show how a rigging map $\eta$ can be defined using 
this choice and that this map is unique (given $\Phi = 
{\cal S}$).  Unfortunately, rigorous results are known to the author 
only when certain additional assumptions are placed on the constraints
(which will be described below), but it is reasonable to conjecture that
similar results hold in the general case.

The result we need for our system is the following:

\vskip4pt plus2pt

\defin{Property A:}
{ There exists a set of generalized states $\langle c,k|$
for $c \in D_C$, $D_C$ an open subset of $\Rl$ containing $0$,
and $k \in D_K$, $D_K$ an open subset of $\Rl^{n-1}$, satisfying
$\langle c,k|\hat{C} = \langle c,k|c$ and $\langle c,k|c',k' \rangle
= \delta (c-c') \delta (k-k')$ and which are complete on the closed
subspace of ${\cal H}_{\rm aux}$  corresponding to the open spectral
interval $D_C$ of $\hat{C}$.  The $\langle c,k|$ are
elements of the (algebraic) dual ${\cal S}^{\rm dual}$ to
${\cal S}$ and the map $F_k: c \mapsto \langle c,k|$ is continuous
with respect to the pointwise convergence topology on
${\cal S}^{\rm dual}$.  Furthermore, the map $F: D_C \times {\cal S}
\rightarrow L^2(D_K,d^{n-1}k)$ given by $F: (c,|\psi\rangle)
\mapsto |\psi\rangle_c$ such that $\langle k|\psi\rangle_c
= \langle c,k| \psi \rangle$ is well-defined and smooth.  }

\vskip4pt plus2pt

Such a result is easy to derive when the constraint is a vector field
with sufficiently regular orbits by simply introducing coordinates
in the space of orbits.  For the case of a Hamiltonian constraint, 
we will need to say something more about the form of the Hamiltonian.
Results are known for the following special cases:

\vskip4pt plus2pt

\item{1.}  The massive free particle: Property A may be checked directly
using the momentum eigenstates.

\item{2.} The so-called separable semi-bound cases (see [17]): It follows
from the integral representation 5.14 of [17] that, when a scattering
operator exists for the `transverse' Hamiltonian $H_1$, there is
a complete set of orthogonal and appropriately normalized generalized
eigenstates $\langle c,k|$ satisfying Property A.
	
\item{3.}     When the constraint is of the form $H = \sum_i p_i^2 + V(q) -E$
and $V\in L^1$:  By extending Lemma IV.28 of [20] from
$C_0^\infty(\Rl^n)$ to ${\cal S}$, Property A reduces to the requirement that
$H$ have purely continuous spectrum.

\vskip4pt plus2pt

Unfortunately, the literature contains less helpful
results than one would like.
This is largely due to the fact that Hamiltonian constraints tend not
to have positive definite kinetic terms, while the literature is 
primarily concerned with the Hamiltonians of  particles moving on a
Riemannian space.  Nevertheless, case 2 above contains nontrivial
cosmological models and 
we suspect that Property A in fact holds in more general situations.
We will therefore assume that our system has Property A without
further justification.

Now, for $|\phi \r \in {\cal S}$, let $\phi(c,k)$
be the function $\l c,k | \psi \r$. 
Using Property A, we can construct the rigging map $\eta_0$ through
$$
(\eta_0 \phi_1)[\phi_2] = \int dc \ \delta(c) \ \int dk \
\phi_1^*(c, k) \phi_2(c,k)
$$ 
which clearly satisfies the criteria of step 5c.  Note that the
action of the delta function is well defined since $\phi_1$ and
$\phi_2$ are continuous in $c$ by property A.

We will now see that this is the unique map (up to an overall scale)
that satisfies 5c.  To do so, consider some generic rigging map $\eta$.
Since $\eta$ must commute
with the constraint, but has only solutions of the constraint in its
image, it is clear that $\eta$ must annihilate the the space $D \subset
{\cal S}$ of states which are in the domain of
${\wh C}^{-1}$ and which are mapped into ${\cal S}$ by 
${\wh C}^{-1}$. This is the space of smooth $\phi(c,k)$ for which
$c^{-1} \phi(c,k)$ is also smooth.  Since any smmoth function that vanishes
at $c=0$ at zero must vanish at least as fast as $c$, this is
in fact the space of all $|\phi \r 
\in {\cal S}$ for which $\phi(0,k) = 0$.  It follows that the kernel of 
$\eta$ includes the kernel of $\eta_0$.

Let us now consider two states $|\phi_1\r, |\phi_2 \r \in {\cal S}$
which are {\it not} annihilated by $\eta_0$; that is, for which
$\phi_1 (0,k)$ and $\phi_2(0,k)$ are nonzero on a positive measure
subset of $D_K$.  Then by 
continuity there is some $\epsilon$ such that
$$ \int dk \ |\phi_i (c,k)|^2 > 0 \ \ \ \ (i = 1,2)$$
for all $|c| < \epsilon$ and such that $[-\epsilon, \epsilon]
\subset D_C$.  We now define $\Pi_{[-\epsilon, \epsilon]}$
to be the projection onto the spectral interval $[-\epsilon, \epsilon]$
of the constraint $\wh C$ and consider the state
$$
|\psi_1 \r = \Pi_{[-\epsilon, \epsilon]} |\phi_1 \r.
$$
Note that $|\psi_1 \r$ and $|\phi_1\r $ map to the same 
element of $\Phi'$ under both $\eta$ and $\eta_0$.
We also define a state
$|\psi_2 \r$ by  the equation
$$
\psi_2 (c,k) = \sqrt{ { {\int dk' |\phi_1(c,k')|^2}   \over   {\int dk'
|\phi_2(c,k')|^2}  }} \phi_2 (c,k)
$$
for $ |c| \leq \epsilon$ and $\phi_2(c,k) = 0$ for $|c| > \epsilon$.
While $|\phi_2 \r$ and $|\psi_2 \r$ map to different elements of 
$\Phi'$,  they map to the same ray in ${\cal H}_{\rm phys}$ under $\eta$
and to the same ray in ${\cal H}_{{\rm phys},0}$ under $\eta_0$.
Note that $\eta_0 |\psi_1 \rangle$ and $\eta_0 |\psi_2\rangle$
have the same norm in ${\cal H}_{{\rm phys},0}$, but are otherwise
arbitrary elements of ${\cal H}_{{\rm phys},0}$.

We will now show that the conditions of step 5c guarantee that
$\eta|\psi_1 \r$ and $\eta |\psi_2 \r$ have the same physical
norm no matter how $\eta$ is defined.  To proceed, 
consider the family $U(\theta)$ of unitary operators
that generate rotations in the two dimensional subspace of $\haux$
spanned by $|\psi_1 \r$ and $|\psi_2 \r$ and note that, for 
fixed $c$, the functions $\psi_i(c,k)$ define elements
$|\psi_{i,c}\r$ of the `transverse' Hilbert space $\h_c \sim L^2(D_K,dk)$.
Such $U(\theta)$
are in fact diagonal in $c$; that is, they satisfy
$$
\l c,k |U(\theta) |\phi \r = \l k | U_c(\theta) | \phi_c \r_c
$$
where the subscripts $c$ on the r.h.s. indicate that the matrix
element is taken in the transverse Hilbert space $\h_c$.
Here, $U_c(\theta)$ is just the unitary operator on $\h_c$
that rotates the subspace
spanned by $|\psi_{1,c} \r$ and $|\psi_{2,c}\r$ and $\l k |$
is the ket for which $\l k |\phi_c \r_c = \phi(c,k)$.  As a result, 
$U(\theta)$ commutes with the constraint $\wh C$ and, since it preserves
the subspace $\Phi$, must belong to the algebra $\bsp$ of observables.
However, this means that it must commute with $\eta$ and define a
unitary operator on the corresponding physical Hilbert space.
It follows that whenever  $(\eta_0 \psi_1)[\psi_1] = (\eta_0 \psi_2)[\psi_2]$, 
we must also have 
$(\eta \psi_1)[\psi_1] = (\eta \psi_2)[\psi_2]$.  Since
$\eta$ provides a positive semidefinite inner product, the functional
$\phi \mapsto (\eta \phi)[\phi]$
in fact defines $\eta$ completely and $\eta$ must be just
$\eta_0$ up to some overall positive scale factor.

\section{4. Superselection Laws}

In contrast with the previous section, the case considered in 
[1] did not result in a unique rigging map.  Instead, a 
large family of maps was found, associated with the existence of certain
`superselection rules.'  It seems a 
reasonable conjecture that, for a
given choice of subspace $\Phi$, the non-uniqueness
of the rigging map is always exactly determined by the 
superselection rules.  While we shall not prove this here, the
discussion
below provides supporting evidence.  Appendix A shows that this is
true for the particular case studied in [1].

Interestingly, the very existence of superselection rules can 
depend on the choice of the dense subspace $\Phi$ of step 5b.
This emphasizes the importance of choosing $\Phi$ based on physical
motivations. Below,
we provide two examples of cases where a
superselection
laws arises: one (in 4.1) in which it seems to come from the `wrong' choice
of $\Phi$, and one (in 4.2) in which its existence reflects
a feature of the classical physics.

{\bf 4.1} {\it The Dependence on $\Phi$.}
For our first example, we will rework the case of section 2 using a 
different choice of $\Phi$.  Property A allows us to introduce a
notion of {\it continuous} states as follows:

\vskip4pt plus2pt

\defin{Definition} {A state $|\phi \r \in \haux$ is said to be
continuous on $\Sigma \subset D_C$ if $\phi(c,k)$ is continuous in
$c$ for each fixed $k$ at every $c \in \Sigma$.}
 
\vskip4pt plus2pt

We will construct $\Phi$ in the following (complicated!) way.  Choose
some interval \break $[-a,a] \subset D_C$.  Now, consider the
subintervals $I_n^- = (-{a \over {2^n}}, - {a \over {2^{n+1}}})$
and $I_n^+ = ({a \over {2^{n+1}}},{a \over {2^n}})$ 
for $n \ge 0$.
Let $R_E$ be the union of the $I^\pm_n$ for even $n$ and $R_O$
be the union for odd $n$.  In addition, consider a family
of projections $\Pi_c$ on $\h_c$ for which the matrix elements 
$\l k| \Pi_c | k' \r_c$ are independent of $c$ and let ${\cal N}_c$
be the subspace of $\h_c$ annihilated by $\Pi_c$.
We now let $\Phi$ be the dense subspace of $\h_{aux}$ containing all
states $|\psi \r$ such that

\vskip4pt plus2pt

\item A) $|\psi \r$ is continuous on $R_E$ and $\lim_{c \rightarrow 0 \
{\rm in} \ R_E}$ exists in $\Co$.
\item B) $|\psi \r$ is continuous on $R_O$ and $\lim_{c \rightarrow 0 \ {\rm
in} \ R_O}$ exists in $\Co$.
\item C) $\Pi_c |\psi_c\r = 0$ at the midpoint of $I^\pm_n$ for each 
odd n.

\vskip4pt plus2pt

The limit in A (B) is taken by considering only
sequences in $R_E$ ($R_O$).

Note that since elements of $\Phi$ are only required to be continuous
separately on the sets $R_E$ and $R_O$, there are now ${\it two}$
natural choices for the rigging map, $\eta_E$ and $\eta_O$:
$$ (\eta_E \phi)[\psi] = \lim_{c \rightarrow 0 \ {\rm in} \ R_E}
\int dk \ \phi^*(c,k) \psi(c,k)$$
$$ (\eta_E \phi)[\psi] = \lim_{c \rightarrow 0 \ {\rm in} \ R_O}
\int dk \ \phi^*(c,k) \psi(c,k)$$

\vskip4pt plus2pt

\remar{Remark\ {1.}\ } {
Note that $\eta_E$ leads to the usual
physical Hilbert space $L^2(D_K,dk)$, whereas $\eta_O$ leads
to a smaller physical space isomorphic to ${\cal N}_c$.}

\vskip4pt plus2pt

\remar{Remark\ {2.}\ } {
For fans of group 
averaging, we mention that the group averaging procedure [1,8,9,12]
does not converge on $\Phi$ (see Appendix B).}

\vskip4pt plus2pt

The existence of these two maps is associated with the following
superselection rule.  Let $\Phi_E \subset \Phi$ contain
those states of $|\psi \r$ for which $\psi(c,k) = 0$ when 
$c \in R_O$ and let  $\Phi_O \subset \Phi$ contain those
for which $\psi(c,k) = 0$ when $c \in R_E$.  Then, for any
$A \in \bsp$, because $[A,C] = 0$, we  have
$\l \phi_E | A | \phi_O \r = 0$ for any $|\phi_E \r \in \Phi_E$, 
$|\phi_O \r \in \Phi_O$.  Such superselection rules then 
descend to the physical level; that is, to the
action of the physical operators on the physical
Hilbert space.

For a general constraint (such as, say,  $p_x = 0$, generating
translation gauge invariance), there is no reason to expect 
superselection rules.  Also, the Hilbert space that results from
$\eta_O$ seems unreasonably small.  Thus, we must regard these features
as artifacts of using the `wrong' choice of $\Phi$.  In contrast, the
physically motivated choice of section 3 produced perfectly satisfactory
results.

{\bf 4.2} {\it Physical superselection laws}
We now turn an example the superselection
rule captures a feature of the corresponding classical
system, and thus appears physically meaningful.  
For this case, we consider systems which differ
slightly from those considered so far.  We now ask only that
our system satisfy `Property B:'

\vskip4pt plus2pt

\defin{Property B:}{  The Hilbert space $\haux$ can be written
as a direct sum $\haux = \h_{disc} \oplus \h_{cont}$
where $\h_{disc}$ is (densely) spanned by {\it normalizable} eigenstates
of $\wh C$ and such that when the system is
restricted to $\h_{cont}$, it
satisfies Property A.}

\vskip4pt plus2pt

Now, let $\Phi = \h_{disc} \oplus \Pi_{cont} {\cal S}$ where
$\Pi_{cont}$ is the projection to ${\cal H}_{cont}$.  Again, there are
two natural choices of rigging map.  First is $\eta_{disc}$,
$$ \eta_{disc} |\psi \r = \l \psi | \Pi_0$$ where $\Pi_0$ is
the projection onto the (normalizable) eigenstates of $\wh C$ with
eigenvalue zero.  Second is $\eta_{cont}$, defined to annihilate
$\h_{disc}$ but otherwise just as in section 3.  Any combination
$a \eta_{disc} + b \eta_{cont}$ for $a,b > 0$ also defines a rigging
map that satisfies the requirements of step 5c.

Again, there is an associated superselection law between
$\h_{disc}$ and $\Pi_{cont} {\cal S}$.  To see this,
note that since $A \in \bsp$ has an adjoint $A^\dagger \in \bsp$, 
we need only show that, for all $A \in \bsp$, $A$ maps
$\h_{disc}$ into $\h_{disc}$ 
and we will be done.  However, since $[A, \wh C] =0$ 
and the domain of 
$A$ contains $\Phi \supset \h_{disc}$, $A$
must map every normalizable eigenvector of $\wh C$ to a normalizable 
eigenvector of $\wh C$ (with the same eigenvalue).  Thus, each $A \in \bsp$ 
preserves $\h_{disc}$, providing us with a superselection rule.
Again, this descends to a superselection rule for the physical
operators on the physical Hilbert space.

However, this time the corresponding {\it classical} system has
a similar feature\footnote{${}^2$}{The argument given below is
an improved version of the one given in Appendix A of [1].}.  To see this, 
recall that when an operator $\wh{A}$ is 
associated with a function A on the classical phase space, 
the discrete eigenvalues of the operator $\wh{A}$ are 
associated with parts of the phase space in which the orbits of
the Hamiltonian vector field of the function A are contained in 
compact regions, while the continuous eigenvalues are associated with
parts of the phase space where these orbits are not contained in 
compact regions.

Suppose then that we have a single classical constraint
$C$. For concreteness, we assume that the phase space $\Gamma$
is a finite dimensional manifold.
Let $\Gamma_{disc}$ be the union of the collection of all orbits $O$
generated by this constraint such that there exists a compact $K_O \in \Gamma$
containing $O$.  We may think of $\Gamma_{disc}$ as the classical
analogue of the space $\h_{disc}$ of 
discrete eigenvectors  of $\wh{C}$.   Let $\Gamma_{cont}$
be the rest of the phase space $\Gamma$.  Now, consider some
function $A$ on the phase space such that $A$ Poisson commutes with $C$.
The exponentiated action of the Hamiltonian vector field defined
by $A$ is a (local) homeomorphism that
maps orbits of $C$ onto orbits of $C$.  Since, for an orbit 
$O \subset \Gamma_{disc}$, every neighborhood $U \subset 
\Gamma$ of $O$ contains some 
compact set $K_U$ which contains $O$, we therefore conclude that 
this exponentiated map cannot take an orbit $O_{disc} \in \Gamma_{disc}$ to an
orbit $O_{cont} \in \Gamma_{cont}$ and vice versa.  Thus, we find that
(in the terminology of [14]) $\Gamma_{disc}$ and $\Gamma_{cont}$ contain
disjoint sets of symplectic leaves of $\Gamma$ and we have a
classical superselection law between the corresponding two parts of the
reduced phase space. 
This seems to be the direct classical analogue of the quantum
superselection rules discussed above; in fact, it is even stronger.
All that is really required in the above argument is that the Poisson
bracket of $A$ and $C$ vanish on the constraint surface.  Thus, 
this superselection rule holds even for the so-called `weak
observables.'

It seems then that we must be careful.  When the space $\Phi$
is chosen to reflect the smooth structure of the phase space, we
have found physically meaningful superselection rules, a reasonable
physical Hilbert space, and a (sufficiently) unique rigging map.
However, when this is not the rationale for choosing $\Phi$, spurious
results may occur.  In the case of the diffeomorphism invariant
states of [1], the corresponding $\Phi$ was chosen to reflect
this structure as it is the appropriate domain of definition for the
operators that were assumed to function as coordinates and momenta.
Thus, within the framework of the auxiliary space of [1] and
modulo questions concerning the Hamiltonian constraint (which was
intentionally ignored), we expect that the superselection
rules of [1] should be taken seriously.

\section{Acknowledgments}

The author would like to thank Abhay Ashtekar, Petr Hajicek,
Atsushi Higuchi, 
Nicholas Landsman, Jurek Lewandowski, 
Jos\`e Mour\~ao, and Thomas Thiemann for many useful discussions.
Special thanks are due to Chris Isham and Karel Kucha\v{r}
for repeatedly asking about the uniqueness of the physical Hilbert
space given by the refined algebraic approach, to Chris
Fewster for especially clarifying discussions and for help in locating
reference [19], and to Carlo Rovelli for discussions on the
significance of the superselection rules.  Finally, many thanks to
Domenico Giulini for pointing out an error in a previous version of
the paper.

\section {Appendix A.  Uniqueness of the construction of Connection
Representation Diffeomorphism Invariant States}

In this appendix, we give a short proof that the rigging maps used
in [1] to solve the diffeomorphism
constraint completely exhaust the set of possible such
maps given the choice of auxiliary space, the definitions of the quantum
constraints, and the dense subspace $\Phi$ chosen in [1].
For a full definition of the terms and notation used below, see
[1].  

Recall from that the auxiliary Hilbert space of [1]
is spanned by a set of orthonormal `spin network states.'  We
shall denote these states by $|\Gamma_{\alpha, k} \r$,
where $\alpha$ is a (piecewise
analytic) graph embedded in a given analytic three manifold and $k$
is an index that takes some finite set of values (this set depends
on the graph $\alpha$).  In addition, (analytic)
diffeomorphisms ${\cal D}$ act on these states by moving the graph
$\alpha$ in the obvious way and permuting the values of the index $k$
allowed by $\alpha$.

The dense subspace $\Phi$ of step 5b 
is the space of so-called smooth cylindrical
functions.  This space contains all finite linear combinations of the
spin network states $\Gamma_{\alpha, k}$ and, for our purposes, 
may in fact be identified with this slightly smaller space.
Following [2], we shall consider only `type I graphs' (see [2]).

As in [1], it is convenient to introduce the subspaces
$\h^{[\tilde \beta]}$ spanned by spin networks $|\Gamma_{\alpha,k}\r$
associated with
graphs $\alpha$ that can be moved by a diffeomorphism to some
graph $\beta$ for which $\tilde \beta$ is the `maximal analytic
extension.'  These subspaces are superselected by the algebra 
$\bsp$ and, on each subspace, there is a corresponding map
$\eta^{[\tilde \beta]}$ defined by:
$$
\eta^{[\tilde{\beta}]} |f\r =
\bigl( \sum_{{\cal D}_1 \in S(\tilde \beta)}
\sum_{[{\cal D}_2] \in GS(\tilde \beta)} {\cal D}_1 {\cal D}_2 |f \r
\bigr)^\dagger
$$ 
where we still need to introduce the set $S (\tilde \beta)$ 
and the quotient space
$GS(\tilde \beta)$.  $S(\tilde \beta)$ is chosen to be any set (and the above
map does not depend on this choice) of diffeomorphisms ${\cal D}_{
\tilde \alpha}$, one for each maximally 
extended analytic graph $\tilde \alpha$
diffeomorphic to $\tilde \beta$, such that ${\cal D}_{\tilde \alpha}$
moves  the extended graph $\tilde \beta$ onto the extended graph $\tilde
\alpha$.  On the other hand, $GS(\tilde \beta)$ (the `graph
symmetry group' of $\tilde \beta$) is the quotient $Iso(\tilde \beta)
/ TA(\tilde \beta)$ where the `isotropy
group' $Iso(\tilde \beta)$ is the group of diffeomorphisms
which map $\tilde \beta$ onto $\tilde \beta$ and the `trivial
action group' $TA(\tilde \beta)$ is the group of diffeomorphisms
that map every edge $e$ in $\tilde \beta$ onto itself.
In the formula above, $[{\cal D}_1]$ denotes the equivalence class of 
${\cal D}_2$ in $GS(\tilde \beta)$.

Any linear combination $\sum_{i \in I} a_i \eta^{[\tilde \beta_i]}$ with
positive coefficients $a_i$ satisfies the requirements of step
5c.  (Note that this sum always converges no matter how big the
coefficients $a_i$ or the index set $I$.)  
We would now like to show that such sums exhaust the set of
all rigging maps.
We will follow the same basic strategy as in the uniqueness proof of
section 3.  That is, we now consider a generic map $\eta$ satisfying 
5c and show that if $\eta^{[\tilde
\beta]} |\phi \r = 0$ for all $\tilde \beta$, then $\eta |\phi\r = 0$
as well. 

Suppose then that 
such that $\eta^{[\tilde \beta]} |\phi_0\r = 0$ for all
$\tilde \beta$.
Since $|\phi_0 \r \in \Phi$, it can be written as a sum of
spin network states.  It will be particularly convenient to 
write it in the form:
$$
|\phi_0 \r  = \sum_{i} \sum_j  c_{ij} {\cal D}_j |\Gamma_{
i} \r
$$
where $c_{ij} \in \Co$, ${\cal D}_j \in {\rm Diff}^\omega$, and
$\{ |\Gamma_{i}\r \}$  is some set of spin network states, 
carefully chosen so that no analytic diffeomorphism maps one 
spin network state in this set onto another. 
Now, it is easily checked that $|\phi_0 \r$ is annihilated by 
the above maps exactly when $\sum_j c_{ij} = 0$ for each $i$.
However, any rigging map that commutes with diffeomorphisms and whose
image contains only diffeomorphism invariant states must also
annihilate states with $\sum_j c_{ij} = 0$.  Thus, $\eta |\phi_0\r = 0$.

Now consider some spin network state $|\Gamma_0\r :=
|\Gamma_{\alpha, k}\r$ such that
$\eta |\Gamma_0\r$ is nonzero (so that $\en |\Gamma_0\r$ is nonzero
as well) and choose any other state
$|\Gamma_1\r \in \h^{[\tilde \alpha]} \cap \Phi$.  We want to construct an
operator $A$ in $\bsp$ that has nonzero matrix elements between
$|\Gamma_0\r$ and $|\Gamma_1\r$.  This can be done by applying just
the kind of `group averaging' that was used in the construction of
$\eta^{[\tilde \beta]}$:
$$
A:= \sum_{{\cal D}_1 \in S(\tilde \alpha)} \sum_{[{\cal D_2}] \in
GS(\tilde \alpha)} 
{\cal D}_1 {\cal D}_2 |\Gamma_1 \r \l \Gamma_0| 
{\cal D}_2^{-1} {\cal D}_1^{-1}. 
$$
This operator is diffeomorphism invariant and finite on $\Phi$ for 
exactly the same reasons as the map $\en$ (and similarly for $A^{\dagger}$).
As a result, it is an element of $\bsp$.

Note that $A|\Gamma_0\r$ is a sum of spin networks that differ from
$|\Gamma_1\r$ only by a diffeomorphism.  Thus, $A|\Gamma_0\r$
maps under $\eta$ to a diffeomorphism invariant state that is proportional
to $\eta |\Gamma_1\r$.  However, the number of terms in this sum is
just the physical norm of the state $|\Gamma_0\r$ as defined through
the map $\en$ (and similarly for $A^{\dagger}|\Gamma_1\r$).  
Let us therefore set 
$N_0 = (\en \Gamma_0)[\Gamma_0]$ and $N_1 = (\en \Gamma_1)[\Gamma_1]$
so that $\eta A^{\dagger} A |\Gamma_0\r
= N_0 \eta A^{\dagger} |\Gamma_1\r =  N_0 N_1 \eta |\Gamma_0\r$.
Applying this distribution to $|\Gamma_0 \r$ we have:
$$
{{ (\eta \Gamma_1)[\Gamma_1]} \over {N_1} } = {{(\eta \Gamma_0)[\Gamma_0]}
\over {N_0}}.$$  As before, this guarantees that when acting on
the subspace $\h^{[\tilde \alpha]} \cap \Phi$, $\eta$ acts just
like $\en$ up to an overall positive scale factor.  Since the
domains of the $\en$'s are orthogonal,
it follows that $\eta$
may in fact be expressed as a sum 
of the $\en$ weighted by positive coefficients.

\section{Appendix B. Convergence of the group averaging procedure}

In this appendix we show that the integral that defines the
group averaged inner product does not (absolutely) converge on 
the entire space $\Phi$ given in the second example of section 4.
Recall that the group averaging proposal [1,8,9,12]
is to introduce the physical inner product
$$
\l \phi, \psi \rp = \int dt \l \phi, e^{i t \wh{C}}
\psi \rp $$
for $\phi, \psi$ in $\Phi$.  If this integrand is in fact $L^1$, then
we may write this as
$$
\lim_{T \rightarrow \infty} \int_{-T}^T \l \phi, e^{i t \wh{C}}
 \psi \r = \lim_{T \rightarrow \infty}
 \l \phi, { {\sin (T\wh{C} )} \over {\wh{C}}} \psi \r
$$
$$
{\hskip 3 cm} = \lim_{T \rightarrow \infty}
 \int_{\Lambda} d \lambda \l \phi (\lambda), 
{ {\sin (T\lambda)} \over {\lambda}} \psi (\lambda) \r_\lambda.
$$
However, we will now show that this limit fails to exist for general
$\phi, \psi \in \Phi$.

For convenience, we assume that $D_C = \Rl$.  Furthermore, we
will take $a= 1$ and introduce the intervals $J_n^-
= (-2^{n+1}, -2^n)$, $J_n^+ = (2^n,2^{n+1})$.  Finally, let
$R'_E = R_E \cup (\cup_{\pm, {\rm even} \ n} J_n^\pm)$,
$R'_O = R_O \cup (\cup_{\pm, {\rm odd} \ n} J_n^\pm)$,
and
let 
$|\psi \r \in \Phi$ be any state such that $\psi (c,k)$
vanishes for $c$ in $R'_O$.

The important 
property of $R'_E$ is that this set is preserved when the
real line is scaled by a factor of $2^k$.  As such, given any
function $\psi(\lambda)$ which is continuous on $R'_E$,
the limit 
$$
\lim_{k \rightarrow \infty}
\int_{R'_E} { {\psi(\lambda) \sin (2^k T \lambda)} \over
{\lambda}}
$$
for large $k$ is just
$$
\psi(0) \int_{R'_E} { { \sin (2^k T \lambda)} \over
{\lambda}} \equiv \psi(0) I(T)
$$
which is independent of $k$.
It follows that the limit
exists for large $T$ if and only if $I(T)$ is constant.

However, we will now show that $I(T)$ is not constant.  Note
that its derivative is
$$dI/dT = \int_{R'_E} \cos(T \lambda) d \lambda$$
and suppose that $T=\pi$.  Then, 
$\int_{J^\pm_n} \cos(\pi \lambda) d\lambda = 0$, but 
we have $\int_{I^\pm_n} \cos(\pi \lambda) >0$ so that $I(\pi) > 0$.
As a result, the group averaging norm does not 
(absolutely) converge for any nontrivial $|\psi \r \in \Phi$
that vanishes on $R'_O$.

\references{}{

\item{[1]} A. \spa{Ashtekar}, J. \spa{Lewandowski}, D. \spa{Marolf},
J. Mour\~ao, and T. \spa{Thiemann}, {\it Quantization of
diffeomorphism invariant theories of connections with local 
degrees of freedom}\/,
J.~Math. Phys. {\bf 36} 6456 (1995); gr-qc/9504018

\item{[2]} A. \spa{Ashtekar}, {\it Non-Perturbative Canonical Gravity},
Lectures notes prepared in collaboration with R.S. Tate, World Scientific,
Singapore, 1991. 

\item{[3]} A. \spa{Ashtekar} and R. S. \spa{Tate}, {\it
An algebraic extension of Dirac quantization: Examples},  
J. Math. Phys. 35 (1994) 6434. 

\item{[4]} B.~\spa{DeWitt}, {\it Quantum Theory of Gravity.I. The
Canonical Theory} Phys. Rev. 160 (1967), 1113

\item{[5]} P.A.M. \spa{Dirac} {\it Lectures on Quantum Mechanics}\/,
Belfer Graduate School of Science, Yeshiva University, 
New York, 1964 

\item{[6]} I.M. \spa{Gel'fand}, N.Ya. \spa{Vilenkin}, {\it 
Generalized Functions: vol. 4, Applications of Harmonic Analysis}, 
Academic Press, New York, London, 1964.

\item{[7]} P. \spa{Hajicek},  {\it Quantization of Systems with
Constraints}
in {\it Canonical Gravity: from
classical to quantum} ed. by
J. \spa{Ehlers}, H. \spa{Friedrich}, Lecture notes in Physics,
Springer-Verlag, Berlin, New York, 1994.

\item{[8]} A. \spa{Higuchi}, {\it Quantum linearization instabilities
of de Sitter spacetime: II}, Class. Quant. Grav. 8  (1991) 1983.

\item{[9]} A. \spa{Higuchi}, {\it 
Linearized quantum gravity in flat space with
toroidal topology},
Class. Quant. Grav. {\bf 8} (1991) 2023.

\item{[10]} C.~\spa{Isham}, {\it Canonical Gravity and the Problem of
Time}, Imperial College, preprint TP/91-92/25, gr-qc/9210011 (1992).

\item{[11]} K. Kucha\v{r}, {\it Time and Interpretations of
Quantum Gravity} in {\it Proceedings of the 4th Canadian
Conference on General Relativity and Relativistic Astrophysics},
ed. G. \spa{Kunstatter}  et. al. 
World Scientific, New Jersey 1992. 

\item{[12]} N. \spa{Landsman}, {\it Rieffel induction as
generalized quantum Marsden-Weinstein reduction}\/,
J. Geom. Phys. 15 (1995) 285-319; hep-th/9305088.

\item{[13]} N. \spa{Landsman} and U. \spa{Wiedemann}, {\it Massless Particles, 
Electromagnetism, and Rieffel Induction}\/,
Rev. Mod. Phys. {\bf 7} 923 (1995); hep-th/9411174.

\item{[14]} N.~\spa{Landsman}, {\it Classical and quantum representation
theory}, in
{\it Proceedings Seminar Mathematical Structures in Field Theory},
E. A. de Kerf and H.G.J. Pijls (eds) (CWI-syllabus, CWI, Amsterdam,
to appear 1995).

\item{[15]} D. \spa{Marolf}, {\it The spectral analysis inner product for
quantum gravity}, preprint gr-qc/9409036, to appear in the
Proceedings of the VIIth Marcel-Grossman Conference, R. Ruffini and
M. Keiser (eds) (World Scientific, Singapore, 1995); D. Marolf,
{\it Green's Bracket Algebras and their Quantization},
Ph.D. Dissertation, The University of Texas at Austin (1992).

\item{[16]} D. \spa{Marolf}, {\it Quantum observables and recollapsing
dynamics}, Class. Quant. Grav. 12 (1995) 1199,
gr-qc/9404053. 

\item{[17]} D.~\spa{Marolf}, {\it
Observables and a Hilbert Space for
Bianchi IX}, Class. Quant. Grav. 12 (1995), 1441; gr-qc/9409049.

\item{[18]}
D. \spa{Marolf} {\it Almost Ideal Clocks in Quantum Cosmology: A Brief
Derivation of Time}, Class. Quant. Grav. {\bf 12} 2469 (1995);
gr-qc/9412016.

\item{[19]} B. \spa{Simon}, {\it Quantum Mechanics for Hamiltonians 
defined as quadratic forms}, Princeton Univ. Press, Princeton, 1971, 
p. 120.

}

\bye